\pdfoutput=1
\documentclass[11pt]{article}

\usepackage{acl}

\usepackage{times}
\usepackage{latexsym}
\usepackage{siunitx} % \num
\usepackage{amsmath}

\usepackage[T1]{fontenc}
\usepackage[utf8]{inputenc}

\usepackage{microtype}

\usepackage{inconsolata}

\usepackage{graphicx}

\usepackage{multirow}
\usepackage{graphicx}
\usepackage{graphics}
\usepackage{caption}
\usepackage{subcaption}  
\usepackage{float}
\usepackage{xcolor}
\usepackage{supertabular}
\usepackage{todonotes}
\usepackage{array}
\title{StockGenChaR: A Study on the Evaluation of Large \\ Vision-Language Models on Stock Chart Captioning}
 \author{Le Qiu \and Emmanuele Chersoni \\
 Department of Language Science and Technology\\
         The Hong Kong Polytechnic University\\
         \texttt{lani.qiu@connect.polyu.hk}, \texttt{emmanuele.chersoni@polyu.edu.hk}}
\begin{document}
\maketitle
\begin{abstract}
Technical analysis in finance, which aims at forecasting price movements in the future by analyzing past market data, relies on the insights that can be gained from the interpretation of stock charts; therefore, non-expert investors could greatly benefit from AI tools that can assist with the captioning of such charts. 

In our work, we introduce a new dataset \textit{StockGenChaR} to evaluate large vision-language models in image captioning with stock charts. The purpose of the proposed task is to  generate informative descriptions of the depicted charts and help to read the sentiment of the market regarding specific stocks, thus providing useful information for investors\footnote{The data and code will be made available on \url{https://github.com/Laniqiu/GenChaR}}. %We present a comprehensive evaluation with Large Vision-Language Models (LVLMs) and we show that XXX...  %Impressed by the success of large language models (LLMs), the study decides to pioneer itself by exploring the capabilities of large vision-language models (LVLMs) on the proposed task. This paper outlines the objectives of the stock captioning task, the dataset we built, and automatic evaluation with some representative general-purpose LVLMs.
\end{abstract}

\section{Introduction}% one page
\begin{figure}[hpt!]
\centering
\includegraphics[width=0.85\linewidth, trim={3.5cm 8cm 0 3.5cm}, clip, keepaspectratio]{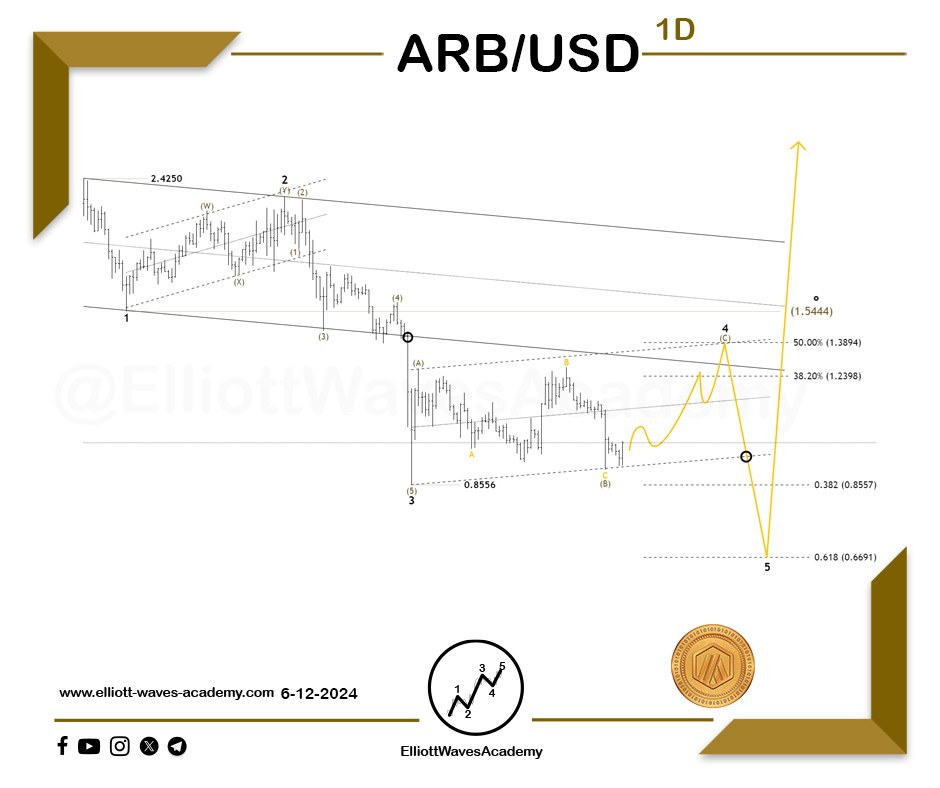}
\caption{Future outlook for ARBUSD on the daily time frame based on Elliott-wave Theory \cite{waves2024future}. The yellow line indicates the predictive movements of the stock.}
\label{fig:analysis_demo}

\end{figure}

In finance, technical analysis is the discipline that aims to predict future price trends based on historical data \cite{ponsi2016technical, edwards2018technical}. Analysts usually anticipate price directions by inspecting \textit{stock charts}, where they are represented in the form of wave patterns (see Figure \ref{fig:analysis_demo}). Such patterns allow traders to identify potential entry or exit positions and make informed investment decisions.
%Stock charts serve as working tools in technical analysis, a discipline that
%aims to predict price trends based on historical data \cite{ponsi2016technical, edwards2018technical}. By utilizing technical analysis techniques, analysts can anticipate the next moves of price movements and reveal them on the charts, also in the form of wave patterns (e.g., Figure \ref{fig:analysis_demo}). These patterns allow traders to identify potential entry or exit positions and make informed investment decisions. Reading these charts, however, pose another challenge. Non-experts can find it difficult to crack the investment suggestions from merely visual cues. 

Reading those charts, however, requires specific financial knowledge and expertise, and it can be challenging for non-expert traders to extract useful insights from merely visual cues. Therefore, we would like to propose framing this problem as an image captioning task where, given an image, an automatic system has to produce a description of its content \citep{vinyals2015show}. In our view, automating the manual interpretation process and producing descriptive texts for the charts would open up a shortcut to understanding the market dynamics for various groups of users, including experienced traders, novices, and individuals seeking support for their fast-paced trading activities.

Previous evaluation work on image captioning largely revolved around general-domain data, with relatively limited coverage of specialized domains \footnote{A summary of evaluation datasets for Image Captioning can be found in Table \ref{tab:datasets_review} in the Appendix.}; and it typically aimed at summarizing the image content with just 1-2 concise sentences \citep{bernardi2016automatic}, which might not be ideal for the goals of technical analysis and financial decision-making. The image captioning setting needs to be adapted for stock chart reading: given an annotated chart image $I$, a system should generate a multi-sentence description $C$ that provides a holistic narrative of the chart, covering the past movements and predictive trends, and ideally with trading advice. The generated text $C$ is expected to be accurate and informative to lead to a well-grounded conclusion for the audience. Additionally, the text should remain as concise and comprehensible as possible, in order to be easily understandable even by less experienced traders.  
\par To this purpose, we introduce \textit{StockGenChaR}, a new dataset for the re-formulated stock-chart captioning task. %Following the recent LLM trends, we first investigated how LLMs can be utilized in this task of stock chart reading. 
To establish baseline performance levels, we tested some representative LVLMs (i.e., LLMs with visual capabilities \citep{li2023evaluating}) by using metrics that focus on different aspects of the generated texts, including n-gram overlap, semantic similarity, sentiment alignment, and accuracy of metadata information. 
\section{Related Work} % half page
\label{sec:lr}
% stock chart 
%As exemplified in Figure \ref{fig:sample_stock_data}, a stock chart can be presented graphically (Figure \ref{fig:sample_stock_data_b}) or numerically (Figure \ref{fig:sample_stock_data_a}). Compared to numerals, graphical charts provide a more intuitive representation of how the price has changed over time.

Prior work around stock charts focused on utilizing the numerical data for goals such as financial return prediction and portfolio optimization \citep{hu2018deep,kusuma2019using,ho2021stock, norasaed2024market}, while other studies made use of the graphical component in image or pattern recognition tasks \cite[e.g.,][]{velay2018stock,zheng2021feature}. However, to our knowledge, the task of stock chart captioning has received limited attention so far.

The most recent approach to the image captioning problem consists of the vision-language pre-training approach (VLP). VLP models are pre-trained on a large amount of image-text pairs, and then fine-tuned for downstream tasks \citep{gan2022vision,chen2023vlp}.  Popular VLP models that can be applied to image captioning tasks include, for example, SimVLM \citep{li2019entangled}, OSCAR \citep{li2020oscar} and CLIP \citep{radford2021learning}. Large Vision Language Models (LVLMs) can be considered as enhanced and ready-to-use versions of VLP models: in recent research work, models such as GPT-4 Vision \citep{openai2023gpt4v}, Gemini \citep{gemini2024demini},  BLIP-2 \citep{li2023blip} and LLaVa \citep{liu2024visual} proved their ability of successfully carrying out several multi-modal tasks, including image captioning and visual reasoning \citep{li2023evaluating, zhang2024mm}.

There have been examples of customized LVLMs that have exhibited some chart reasoning abilities. For example, \citet{liu2023mmc} developed MMCA, a MultiModal Chart Assistant achieving state-of-the-art performance on several chart question answering benchmarks; they also introduced a new and more challenging benchmark with nine different tasks evaluating reasoning capabilities over charts and plots, and reported that even the most sophisticated LVLMs have important limitations in interpreting charts. The works of \citet{bhatia2024fintral} and \citet{xie2024open} both introduced large instruction datasets for tuning LVLMs for the financial domain, together with two models, FinTral and FinLLaVA, that excel in solving tasks related to the interpretation of financial tables and charts. 

Although such works challenge models in visual question answering on charts, we believe that framing stock chart understanding as an image captioning task would be closer to the needs of investors and practitioner in the financial industry, as image captioning could provide trend interpretations beyond the constrained setting of question-based benchmarks. To our knowledge, this type of task is not covered by any of the existing benchmarks for financial chart understanding.

In our study, we aim at filling this gap by building a new dataset for stock chart captioning, \textit{StockGenChaR}.  We will also present a systematic evaluation of the most commonly used LVLMs on the new benchmark.

\section{Dataset Creation}
\label{sec:data}
\begin{figure}[htb!]
  \centering
  \includegraphics[trim={0 5cm 0 1.8cm}, clip, keepaspectratio,scale=0.2]{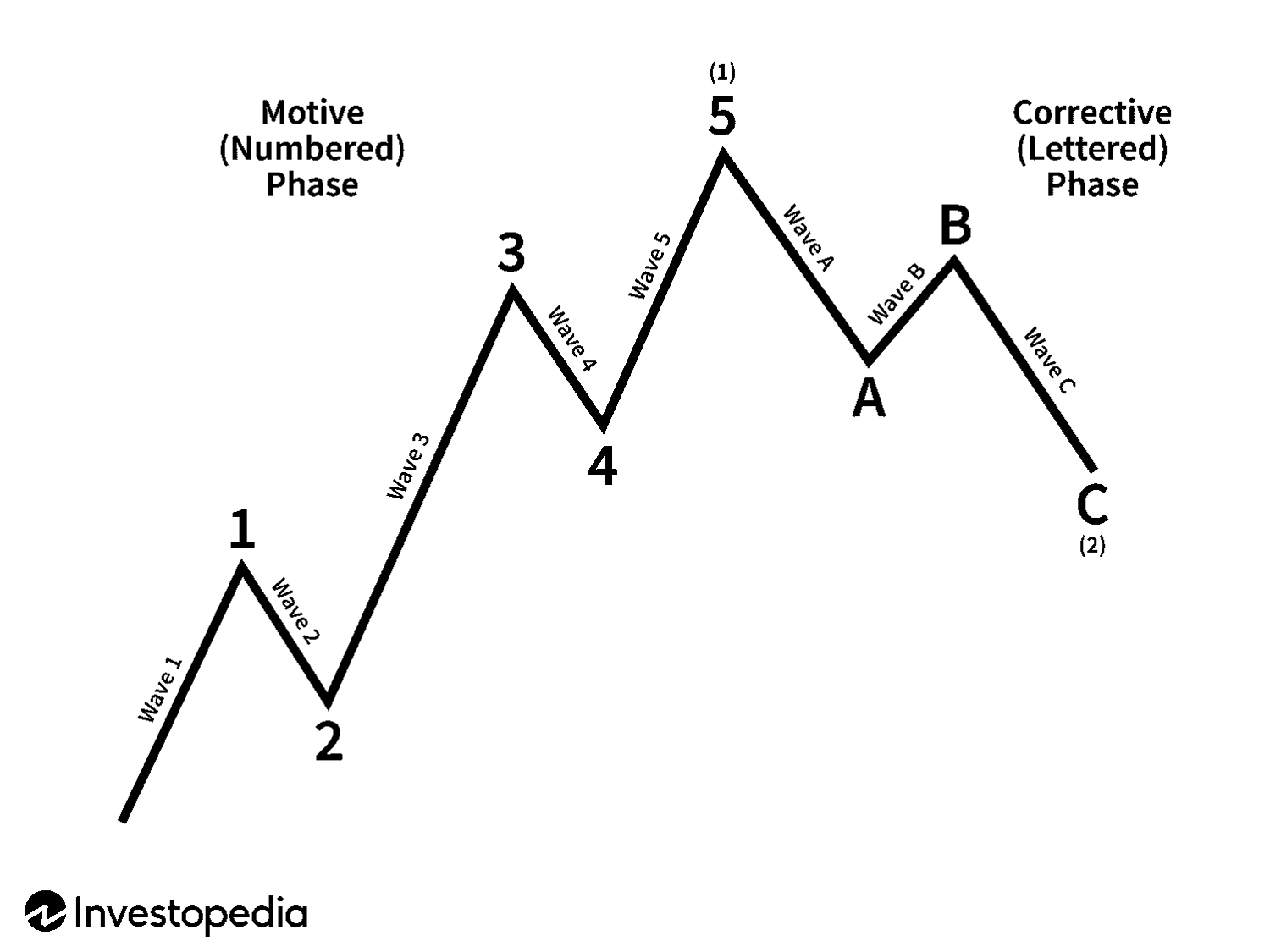}
  \caption{An EWP cycle.\protect\footnotemark}
  \label{fig:ewp}
  
\end{figure}
\footnotetext{Source: \url{https://www.investopedia.com/terms/e/elliottwavetheory.asp}}

\subsection{Sample Collection}
\par We identified an ideal source for this chart captioning task: the \textit{ElliottWave-Forecast} website \footnote{\url{https://elliottwave-forecast.com}}. ElliotWave-Forecast is a worldwide top-notch technical analysis company, providing a wide range of coverage across about 80 markets, including Forex, Commodities, World Indices, and U.S. stocks \& ETFs \citep{ew2024au}. The analyst team uses Elliott Wave Principle (EWP) as a major tool for chart analysis and offers forecasting and instructive guidance to its clients. EWP is a popular technical analysis approach: it is based on the belief that market prices have a tendency to move infinitely in a cycle (see Figure \ref{fig:ewp}) in all time frames, exhibiting repetitive wave patterns \citep{poser2003applying}. EWP provides the theoretical foundations for chart analysis and for the automatic completion of chart patterns within a specific timescale. The use of EWP also makes it easier to understand the charts and the captions, as it annotates the waves with the so-called \textit{degrees} (the alphabetical or numeral indices along the wave patterns in Figure \ref{fig:ewp}). 

\begin{figure}[htb!]
\centering
\small
\begin{subfigure}{0.3\linewidth}
\includegraphics[width=\linewidth, keepaspectratio]{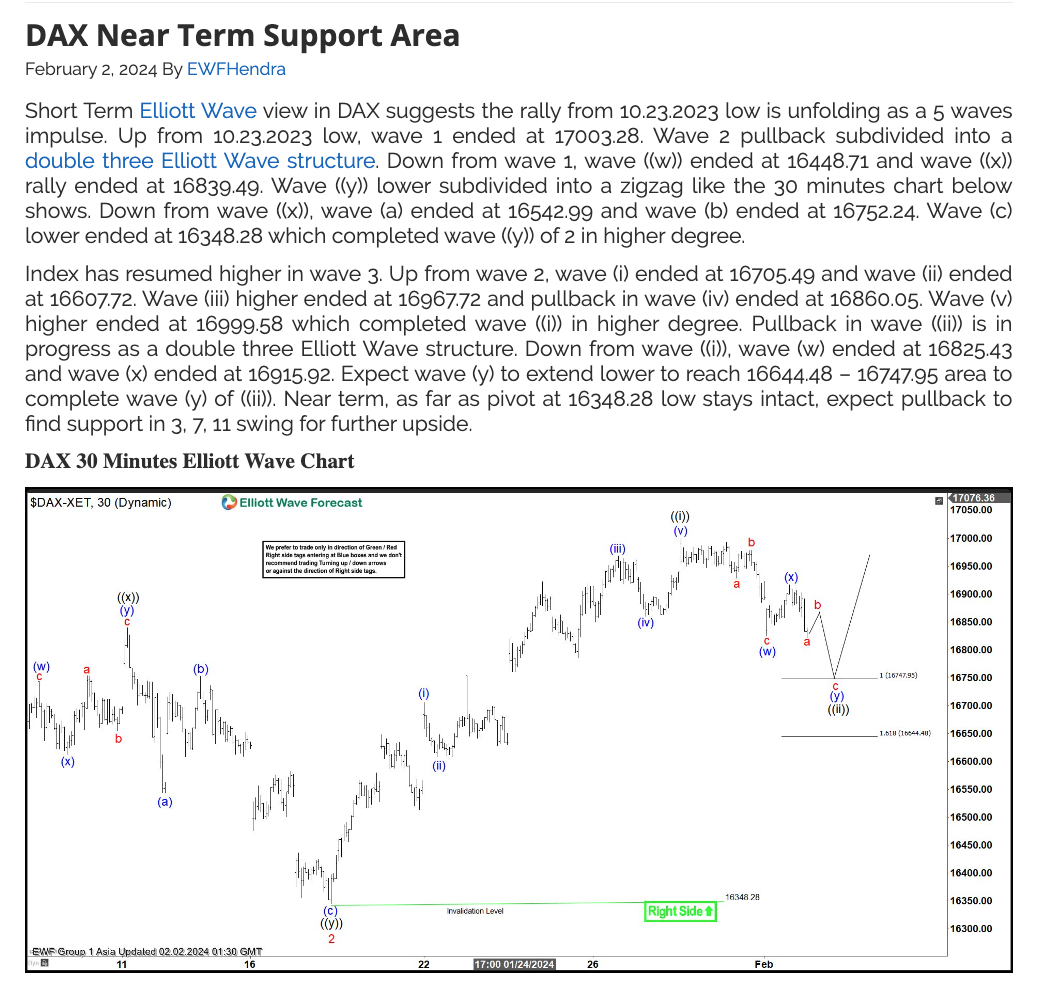}
\caption{Article}
\label{fig:article-demo-a}
\vspace{1.5cm}
\end{subfigure}
\begin{subfigure}{0.68\linewidth}
\begin{subfigure}{\linewidth}
  \includegraphics[width=\linewidth, trim={0 19.5cm 0 2.8cm}, clip, keepaspectratio]{fig/article.png}
  \caption{Textual content}
  \label{fig:article-demo-b}
\end{subfigure}
\begin{subfigure}{\linewidth}\centering
  \includegraphics[width=0.8\linewidth, trim={0 0 0 17cm}, clip, keepaspectratio]{fig/article.png}
  \caption{Chart image}
  \label{fig:article-demo-c}
\end{subfigure}
\end{subfigure}
\caption{A sample article from ElliottWave-Forecast in Fig. \ref{fig:article-demo-a}, with the image and text displayed separately \citep{ewf2024news}. In the image captioning task, Figure \ref{fig:article-demo-c} is the image input \textit{I}, and Figure \ref{fig:article-demo-b} the \textit{gold} caption \textit{C}.}
\label{fig:article}

\end{figure}
\par The analysts publish their analysis (in English) together with chart images (see Figure \ref{fig:article-demo-a}). For our dataset, we downloaded the articles released on the website by February, 2024. Each article was split into images and texts. We removed images that are not target charts, and texts and mark-up that are unnecessary, such as authorship, HTML tags, advertisements, and so on. So far, we only kept the samples containing one single stock chart (some articles may include two or more), to ensure a collection of one-to-one rather than one-to-many image-caption pairs. In addition, samples in which the text body is too long ($>$ 400 words) or too short ($<$100 words) have been excluded, according to the statistics on text length. The remaining stock chart-text pairs are our final dataset items, for a total of \num{1972} chart-caption pairs. %The dataset is randomly split into a train set and test set with a 8:2 ratio. We then used the test set for the evaluation described in section \ref{sec:exp}. \todo{EMM: now we can comment this out, right? Now the evaluation takes the entire dataset into account}

\subsection{Chart Annotation}
\par As shown in Figure \ref{fig:article-demo-c}, stock charts map the price on the Y-axis against the time on the X-axis, and they typically come with several annotations. We categorized such annotations in four main types: Degree, Time, Price, and Add-on. \textit{Degree} refers, roughly speaking, to a price movement; \textit{Add-on} includes the information that is additionally applied to the charts, such as reminder messages and titles, while \textit{Time} and \textit{Price} are self-explanatory. We also annotated the endpoints of predictive patterns (categorized as \textit{Point}) for further studies on automatic pattern completion\footnote{We did not annotate the endpoints for historical patterns, because they can be automatically generated with price data.}. The taxonomy and descriptions of the annotations are presented in Table \ref{tab:tax}. %Among them, elements including \textit{Degree}, \textit{Time} and \textit{Price} are the \textit{metadata} that is directly related to and most essential for chart reading, while \textit{Add-on} includes the information that is additionally applied to the charts, such as reminder messages and titles. We also annotated the endpoints of predictive patterns (categorized as \textit{Point}) for further studies on automatic pattern completion\footnote{We didn't annotate the endpoints for historical patterns, because they can be automatically generated with price data.}. The taxonomy and descriptions of the annotations are presented in Table \ref{tab:tax}.
\begin{table}[hbt!]
\centering
\caption{A taxonomy of charting annotations. Here OHLC is used as a general term for \textit{OHLC}, \textit{Adj.} and \textit{Volume} data. OHLC stands for Opening, Highest, Lowest, and Closing prices of a financial instrument during a timeframe, while \textit{Adj.} is the adjusted closing price accounting for corporate actions, and \textit{Volume} refers to the transaction amount.}
\label{tab:tax}
\begin{tabular}{p{0.2\linewidth}p{0.7\linewidth}}
  \hline
Category & Description\\
\hline
Degree & EWP degrees \\
Time & X-axis ticks, timestamp, time markers\\
Price & Y-axis ticks, OHLC, price markers\\
Add-on & Supplementary indicators, annotations, and watermarks, etc. \\
Point & Endpoints of the predictive patterns \\
\hline
\end{tabular}
\end{table}
\section{Evaluation with LVLMs}
\label{sec:exp}
% \subsection{Setting}
\label{sec:eval_set}
\subsection{Model Choice}
% models, metrics+ qa, 
We ran evaluations with five recent general-purpose LVLMs that have showcased impressive capabilities in image captioning and visual question answering tasks \citep{li2023evaluating, zhang2024mm}, including GPT-4V \citep{openai2023gpt4v}, mPLUG-Owl2 \citep{ye2023mplug2}, LLaVA \citep{liu2024visual}, Instruct-BLIP \citep{dai2024instructblip} and Gemini \citep{gemini2024demini}. Each LVLM was prompted with the instruction below to produce candidate captions, and evaluated in a zero-shot setting. \footnote{We had a pilot study with a few samples and some candidate instructions. The presented instruction achieved good performance overall. We observed that the few-shot learning setting is not applicable given the nature of the presented task, and thus we did not conduct the few-shot experiments}
\begin{itemize}
    \item Instruction: \textit{Based on the chart image, generate a text around 100 to 400 words, describing the historical price movements and predictions and concluding the opinion of the chartist towards the stock trends.}
\end{itemize}
% metrics
\subsection{Evaluation Metrics}
\subsubsection{Text Similarity Metrics}
\par Regarding evaluation metrics, we have considered the most popular ones, which are mainly based on n-gram overlapping, including BLEU \citep{papineni2002bleu}, ROUGE \citep{lin-2004-rouge}, METEOR \citep{banerjee2005meteor} and CIDEr \citep{vedantam2015cider}. %and SPICE \cite{anderson2016spice}. 
We also included two semantic-oriented metrics that are based on contextualized embeddings: BERTScore \citep{zhang2019bertscore}, and paragraph-level cosine similarity, denoted as COS$_F$. After a pilot study, we selected the OpenAI embedding model as the best one to measure COS$_F$ (see section \ref{sec:eval} in the Appendix for more details).
% dimension： general textual 
\subsubsection{Fine-grained Examination}
\par Additionally, observing that the metrics above measure textual similarity in general, we also attempted to have a closer examination of the generated captions. To this purpose, we further analyzed the results in terms of \textit{SA} and \textit{IoU}. 
\\\textbf{IoU  (Intersection over Union)}. We borrowed this metric from the field of object detection to measure how much important information has been included in the candidates relative to the references. In object detection, it measures the overlap over the union of predicted and ground-truth objects \cite{everingham2010pascal}. Different from its original formulation, an adaption has been made to fit into the settings of our work. In object detection, not all objects present in an image are annotated as ground truth. Also, a predicted object is only considered correct if it matches a labeled ground truth object.  
\par In contrast, we assume that all metadata items present in the image are potentially valid and relevant for narration. Therefore, a candidate text should be rewarded whenever it correctly mentions an item from the image's full annotation --- even if that item is not found in the reference text. Such \textit{extra} information is still accurate and should neither be penalized nor neglected. Meanwhile, we attempt to bridge between the candidate and the reference. For this, we measure how the candidate text covers the objects compared to the reference text, while using the annotations as a \textit{background} that has all possible metadata items. Let $O_{a}$ be all metadata appeared in the image, and $O_{c}$ and $O_{r}$ be the metadata that is mentioned in the candidate and reference texts respectively. For each candidate-reference text pair $(c, r)$, we first count how many objects they mention that also appear in $O_{a}$:
\[
  Hit_{c} = |O_{c} \cap O_{a}|,  \\
  Hit_{r} = |O_{r} \cap O_{a}|
\]
Then the relative coverage of $c$ to $r$ can be a ratio: 
\[
  \frac{Hit_{c}}{|O_{a}|} / \frac{Hit_{r}}{|O_{a}|}
\]
To further account for the impact of text length and large values, we redefine the IoU formula as:
formulate IoU as:
\begin{equation} 
  \label{equa:iou}
  IoU = \log\left(1 + \frac{\frac{|O_c \cap O_a|}{L_c}}{\frac{|O_r \cap O_a|}{L_r} + \epsilon}\right),
\end{equation}
where $L_c$ and $Lr$ is the text length of $c$ and $r$ respectively, $\epsilon$ is a constant value to prevent division by zero and a logarithm could smooth the results.\footnote{$\epsilon$ takes the value of \num{1e-6} during calculation.}
\par The dataset contains three categories of metadata  i.e., the \textit{metadata}, including \textit{Degree}, \textit{Time} and \textit{Price}. For each sample, the IoU score is an aggregated result across these three categories. The IoU value would always be non-negative. A higher value may suggest that a more fine-grained description has been given, with more important information or meta information has been referred to. Conversely, lower scores may indicate under-description, omission of key elements. 
\\\textbf{SA (Sentiment Alignment)}.
The reference text states the historical movement of price, then outlines a prediction for future trends, indicating the existence of opinions, i.e., sentiment. An appropriate candidate is expected to express a similar sentiment in general. Accordingly, sentiment analysis can be performed on the reference and candidate texts to evaluate whether their sentiments were aligned. The assessment of sentiment alignment is presented as a text classification problem on three polarity categories, including \textit{positive}, \textit{negative} and \textit{neutral}. The SA score of each reference-candidate pair is formulated as below:
\[
\text{SA}(x, \hat{x}) = 
\begin{cases}
1, & \text{if } x = \hat{x} \\
0, & \text{otherwise}
\end{cases},
\]
where $x$ and $\hat{x}$ denote the reference and candidate text, and $y$ and $\hat{y}$ denote their respective sentiments, which can be obtained using a BERT model fine-tuned for financial sentiment analysis \footnote{\url{https://huggingface.co/ahmedrachid/FinancialBERT-Sentiment-Analysis}}, 1 indicates a correct sentiment alignment between $x$ and $\hat{x}$  and 0 a mismatch.
\par Importantly, we don't perform sentiment analysis over the entire text. What is valued the most is the portion that contains opinions towards the future, which are typically located in the final part of the text. Therefore, our analysis only focus on the last a few sentences of a text. Given that this is an open-ended generation task and sentence-level alignment between reference and candidate texts seems infeasible, we measure aggregated sentiments: The sentiment of each text segment is computed from the predicted labels of its individual sentences through a weighted voting approach, in which sentences closer to the end are assigned higher weights. 
\begin{table*}[htb!]
  \centering
  \caption{A summary of the evaluation results on the dataset (\num{1972} instances). Several images are rejected by GPT-4V, due to sensitivity concerns or other reasons such as image quality, according to the feedback from the model. The \textit{Valid} column contains scores for valid captions only (1915 instances). Out of 1972 GPT-4V responses, 57 are found invalid. 4-gram scores are reported for BLEU and CIDEr, and F-score for BERTScore and for SA. The best scores are indicated in \textbf{bold} among all samples or \underline{underlined} among valid samples. For all metrics, scores are reported in percentage (\textbf{\%}) and rounded half up to two decimal precision and higher values indicate better performance.}
  \label{tab:res-eval}
  \begin{tabular}{l c c c c c c}
      \hline
       & \multicolumn{2}{c}{\textbf{GPT-4V}} & \multirow{2}{*}{\textbf{mPLUG-Owl2}}  & \multirow{2}{*}{\textbf{LLaVA}} & \multirow{2}{*}{\textbf{InstructBLIP}} &  \multirow{2}{*}{\textbf{Gemini}} \\
       & Overall & Valid & \\
      \hline
      \textbf{BLEU} & .65 & .66 & .44 & .49 & .25 & \textbf{0.73} \\
      \textbf{CIDEr} & .58 & .60 & .26 & .24 & .05 & \textbf{0.62} \\
      \textbf{METEOR} & \textbf{15.32} & \underline{15.36} & 11.05 & 11.24 & 7.14 & 15.05 \\
      \textbf{ROUGE$_L$} & 11.74 & 11.78 & 11.39 & 11.28 & 10.07 & \textbf{12.20} \\
          % SPICE & 8.52 & & &\\
          \hline
      \textbf{BERTScore} & \textbf{81.91} & \underline{81.94} & 80.22 & 79.51 & 77.63 & 81.71 \\
      \textbf{COS$_F$} & 71.10 & 71.46 & 49.56 & 41.16 & 36.77 & \textbf{73.32} \\
          \hline
      % \textbf{SA} & \textbf{51.36} & 51.22 & 44.83 & 47.31 & 41.84 & 46.86 \\
      \textbf{SA} & \textbf{38.04} & 37.74 & 35.04 & 31.73 & 32.89 & 35.58 \\
      \textbf{IoU} & 30.88 & 31.31 & 14.94 & 5.19 & .43 & \textbf{38.15}\\
 
      % \textbf{IoU} & 44.98 & 45.16 & 11.64 & 3.69 & .26 & \textbf{56.98}\\
      \hline
  \end{tabular}
\end{table*}
\begin{figure}[htb!]
  \centering
  \includegraphics[width=0.9\linewidth, trim={0 0cm 0 0cm}, clip]{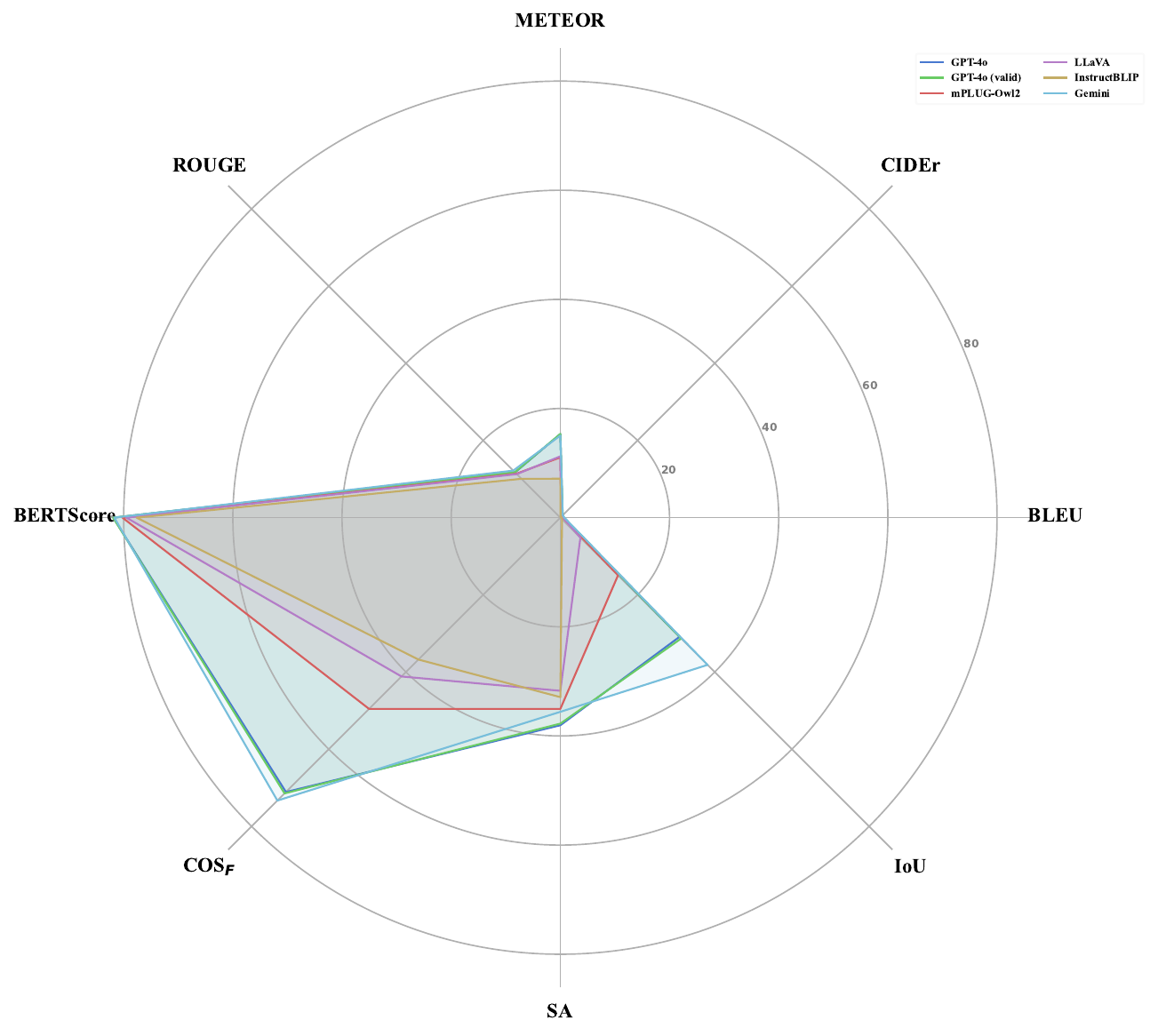}
  \caption{Radar chart of all evaluation scores. All values are presented on a 100\% scale.}
  \label{fig:radar-all}
\end{figure}
\section{Results and Discussion} 
\label{sec:res}
Table \ref{tab:res-eval} presents the zero-shot evaluation results.
\par The instant observation is that all the LVLMs, despite their excellence in image understanding tasks, have achieved very low values for the evaluation metrics based on n-gram overlap (roughly the upper part of Table \ref{tab:res-eval}). The radar chart in Figure \ref{fig:radar-all} presents the extremely unbalanced distribution between these common metrics and those introduced by us. Standard evaluation metrics for image captioning tasks --- BLEU, ROUGE, CIDE$_r$ and METEOR --- consistently report low scores for all models, even for the top ones like Gemini and GPT-4V. This is actually not unexpected: given the domain specificity and the length of the reference texts, it would have been surprising - and possibly an index of data contamination - to observe high values for such metrics. 
\par For SA and IoU, GPT-4V and Gemini outperform the other models. It should be noticed that Gemini scores are much better on the IoU metric, meaning that its responses are more likely to contain more of the correct chart metadata, and has better coverage. In general, LVLMs achieve higher scores in BERTScore and COS$_F$, showing that even if they do not use the same words (i.e. low n-gram overlap), the texts generated by some of the models have high conceptual similarity with the reference. On the other hand, the SA scores, reflecting the alignment between model outputs and opinions of the analysts, have only around 40\% in the best models. Higher values would be desirable, as this metric might be the most relevant in affecting the stock buying decisions of the investors and we would like LVLMs to be as much aligned as possible with human experts' insights, in order for them to be deployed in real-world applications. 
\par Examples are provided in Table \ref{tab:text_comp} (in the Appendix). From them, we can find that the reference text and model output appear to differ in narration. In this sense, the low scores are quite understandable, as these metrics in nature rely on lexical overlapping. While these methods may perform well on regular datasets, where texts are in short forms and show limited possibilities for paraphrasing, their effectiveness can be predictably compromised in this context which presents a open-ended long text generation task.  The extremely low values of BLEU or CIDEr suggest they could be inappropriate for the evaluation of this task. Nonetheless, despite their variance in wording and constructions, some model outputs --- such as those from Gemini and GPT-4V as listed --- remain faithful to the given chart and provide meaningful interpretations for readers. Therefore, relying solely on traditional metrics would seem unfair. In this sense, the use of alternative metrics --- BERTScore and COS$_F$ seem inevitable. BERTScore, according to its formulation (please refer to \citet{zhang2019bertscore} for details), measures lexical similarity using contextualized embeddings from Transformers, which is more flexible and effective compared to its precedents. COS$_F$, on the other hand, encodes the entire text as vectors and then measures semantic similarity in a more blunt manner. 
\par Additionally, we observe that the reference texts which are produced by professional analysts not only uncover the future price trends, but also try to disclose the full picture of the movement, i.e., to provide detailed context and justification for their description or analysis, making it more reliable. During the process, EWP markers (i.e., the numerical and alphabetical annotations around the waves) are frequently referred to, which in some way enhances the accessibility of the text. In contrast, the performance of LLMs varies significantly. Strong models like Gemini and GPT-4V appear to be aware of the EWP theory and able to recognize the visual elements such as markers and labels on the chart. In contrast, weak models could even fail in recognition. For example, the red-highlighted sentence in Table \ref{tab:text_comp} indicates that the LLAVA model could not recognize the time frame of the chart. From its generated textual description, nor can we find traces that the model understands the EWP theory or master the skills of giving suggestions for investment. Text-similarity metrics such as BERTScore, ROUGE, may fail to capture these subtle distinction in their measurement, and that is the reason that IoU is introduced in the evaluation. 
% iou
\begin{table}[hpt!]
    \centering
    \caption{A breakdown of the IoU scores. The reported values (in \%) represent the metadata coverage within different categories. For instance, suppose $n$ Time markers are included in the full annotation and $m$ found in the reference text, the \textit{Time} coverage of the reference is then calculated as $\frac{m}{n}$. }
    \label{tab:iou}
    % \small
    \begin{tabular}{p{0.35\linewidth} >{\centering\arraybackslash}p{0.12\linewidth} >{\centering\arraybackslash}p{0.12\linewidth} >{\centering\arraybackslash}p{0.12\linewidth} }
    \hline
    & \textbf{Degree} & \textbf{Time} & \textbf{Price} \\
    \hline
    \textbf{Reference }&  26.83 & 20.09 & 4.65\\
    \hline
    \textbf{GPT-4V }       & 3.72 & 7.07 & 5.83\\
    \textbf{GPT-4V (valid)}& 3.69 & 7.16 & 5.86 \\
    \textbf{mPLUG-Owl2 }   & .26  & 4.79 & .22 \\
    \textbf{LLaVA }        & .01  &  1.58& .05\\
    \textbf{InstructBLIP } & .00\protect\footnotemark  &  .13 & .01\\
    \textbf{Gemini }       & 5.47 & 12.01& 5.79 \\
    \hline
    \end{tabular}
\end{table}
\footnotetext{The actual value is above zero, but it is displayed as zero due to standard rounding.}
\par Besides from the overall scores presented in Table \ref{tab:res-eval}, Table \ref{tab:iou} provides more statistics around IoU. These objective and quantitative results support our observation that human chartists tend to favor including metadata especially wave degrees, while LLMs are less attentive in their generation. 
\par Regarding SA, we reported its F1-scores in Table \ref{tab:res-eval}. The confusion matrices of sentiment alignment  are revealed in Figure \ref{fig:matrix_sa} in the Appendix. A majority of errors occur between the \textit{neutral} class and the others. Although the sentiment analysis model could be to blame, we noticed that subtle mismatches between the narrative tones could have contributed to the low scores. For instance, as shown in Table \ref{tab:sa_case} (in Appendix), the reference text expresses a positive outlook towards the stock, while Gemini, although agrees, adopts a more cautious and restrained tone. 
\par Based on these findings, it is evident that the some LLMs could have posed the capability of stock chart interpretation and demonstrate a certain level of financial domain knowledge, even in cases where their overall performance is not satisfying. %However, for practical deployment in financial scenarios, task-specific fine-tuning is still required. Also, given the sensitive nature of financial data, directly using open-sourced model may pose risks related to confidentiality or information breach. Also, despite our efforts on evaluation metrics, automatic evaluation metrics alone may be insufficient to capture all aspects of the text quality, especially in our setting, where the task involves generating long-form texts intended for human readers, including amateurs to finance and stocking. In this sense, further explorations with customized approaches and the inclusion of human evaluation seem necessary. On the bright side, we have seen that these LVLMs, close-sourced or open-sourced, are equipped with the visual recognition and textual generation ability. It is therefore safe to assume that the performance on the presented task can be further improved via a more systematic search for optimal prompt instructions, and possibly by introducing customized architectures.
\par In the future, we believe that the performance can be further improved via a more systematic search for optimal prompt instructions, and possibly by introducing customized architectures for the task.

\section{Conclusion}
\label{sec:conclusion}

In this work, we have introduced StockGenChaR, a dataset for stock chart captioning, and we have reported the preliminary work on the captioning task, including the creation of the benchmark and a preliminary evaluation with some popular LVLMs.  By first exploring these LVLMs, we hope to find out their capacity in this stock chart captioning without additional fine-tuning, and also to identify proper evaluation methods for the task. 
\par The current findings suggest that these LVLMs could have possessed limited capabilities of stock chart captioning. However, for practical deployment in financial scenarios, task-specific fine-tuning is still required. Also, given the sensitive nature of financial data, directly using open-sourced models may pose risks related to confidentiality or information breach. Also, despite our efforts on evaluation metrics, automatic evaluation metrics alone may be insufficient to capture all aspects of the text quality, especially in our setting, where the task involves generating long-form texts intended for human readers, including amateurs to finance and stocking. In this sense, further explorations with customized approaches and the inclusion of human evaluation seem necessary. On the bright side, we have seen that these LVLMs, close-sourced or open-sourced, are equipped with the visual recognition and textual generation ability. It is therefore safe to assume that the performance on the presented task can be further improved via a more systematic search for optimal prompt instructions, and possibly by introducing customized architectures.

\section*{Limitations}
% limitation and future work
Our work has some limitations that have to be acknowledged. 
First, the paper presents only preliminary evaluation results with general-purpose LVLMs. %while there are more VLP models and customized LVLMs to be explored (e.g., those mentioned in Section \ref{sec:lr}). 
For future work, we plan to experiment with more customized LVLM architectures (e.g., those mentioned in Section \ref{sec:lr}) to further push the boundaries of model performance on the proposed chart captioning task.
\par Also, the present work only adopts automatic metrics for evaluation. From Table \ref{tab:res-eval}, we can find that metrics that are based on word overlapping such as BLEU and CIDEr, are insufficient in evaluating long texts; the two embedding-based methods, BERTScore and COS$_F$, despite capturing semantic similarity, demand further examination because they could fail to measure the user \textit{accessibility} of the generated texts, as the task requires. In the future, we plan to explore more sophisticated prompting strategies to explicitly target accessibility and other \textit{desiderata} aspects of the generated captions, and to 
include human annotators for evaluating the generated texts. 
% \par For future human evaluation, we decide to introduce 4 fine-grained dimension for  fluency, adequacy, faithfulness and accessibility, with reference to previous studies such as \cite{howcroft2020twenty, celikyilmaz2020evaluation}.
\section*{Acknowledgements}

The authors acknowledge the support from the project “Analyzing the semantics of Transformers representations for financial natural language processing” (ZVYU), funded by the Faculty of Humanities of the Hong Kong Polytechnic University (PolyU-UGC). We would also like to thank our reviewers for the constructive feedback.

\section*{Ethical Considerations}
% risk
\par The models and data used in this study are intended for research purposes only. No personally identifiable information or sensitive content is involved. Therefore, we believe the ethical risk of this work is minimal.
% term of use or license
\par We utilized several models in this study, each governed by its individual license --- LLaVA is released under the Apache License 2.0, mPLUG-Owl2 under the MIT License, and InstructBLIP under the CC BY-NC 4.0 License. For other models and the data we collected without publicly specified licenses, we used them in accordance with the terms of service or usage guidelines by their original provider, where available. Additionally, ChatGPT was employed as a writing assistant under API terms for translation and grammar checking purposes.
\par The annotators were recruited online from Mainland China. All had undergraduate or master’s degrees in computer-related disciplines and were employed in relevant industries. Only individuals who passed a qualification test were selected to participate in the annotation task. All annotators provided informed consent, received compensation on a per-annotation basis in accordance with local labor standards, and retained the right to withdraw from the study at any time.
% ethnical approval
\par This study has been approved by the Institutional Review Board (IRB) from the Department of Language Science and Technology of the Hong Kong Polytechnic University. 
% \newpage
\section*{Acknowledgement of Data Usage}
The data used in this research was obtained from Elliott Wave Forecast with the necessary permission for usage. The provider has explicitly granted consent, ensuring compliance with relevant legal, ethical, and regulatory requirements. We affirm that the data will be handled responsibly and utilized strictly within the agreed scope.
\bibliography{ref}
\appendix   % after refs
% \newpage

\renewcommand{\arraystretch}{1.5} 
\section{Embedding Model Selection}
\label{sec:eval}
\par
\begin{figure*}[t!]
  \centering
  % \begin{subfigure}{0.28\linewidth}
  \begin{subfigure}{0.45\linewidth}
    \centering       
  \includegraphics[width=\linewidth, trim={0 0  8.5cm 0}, clip]{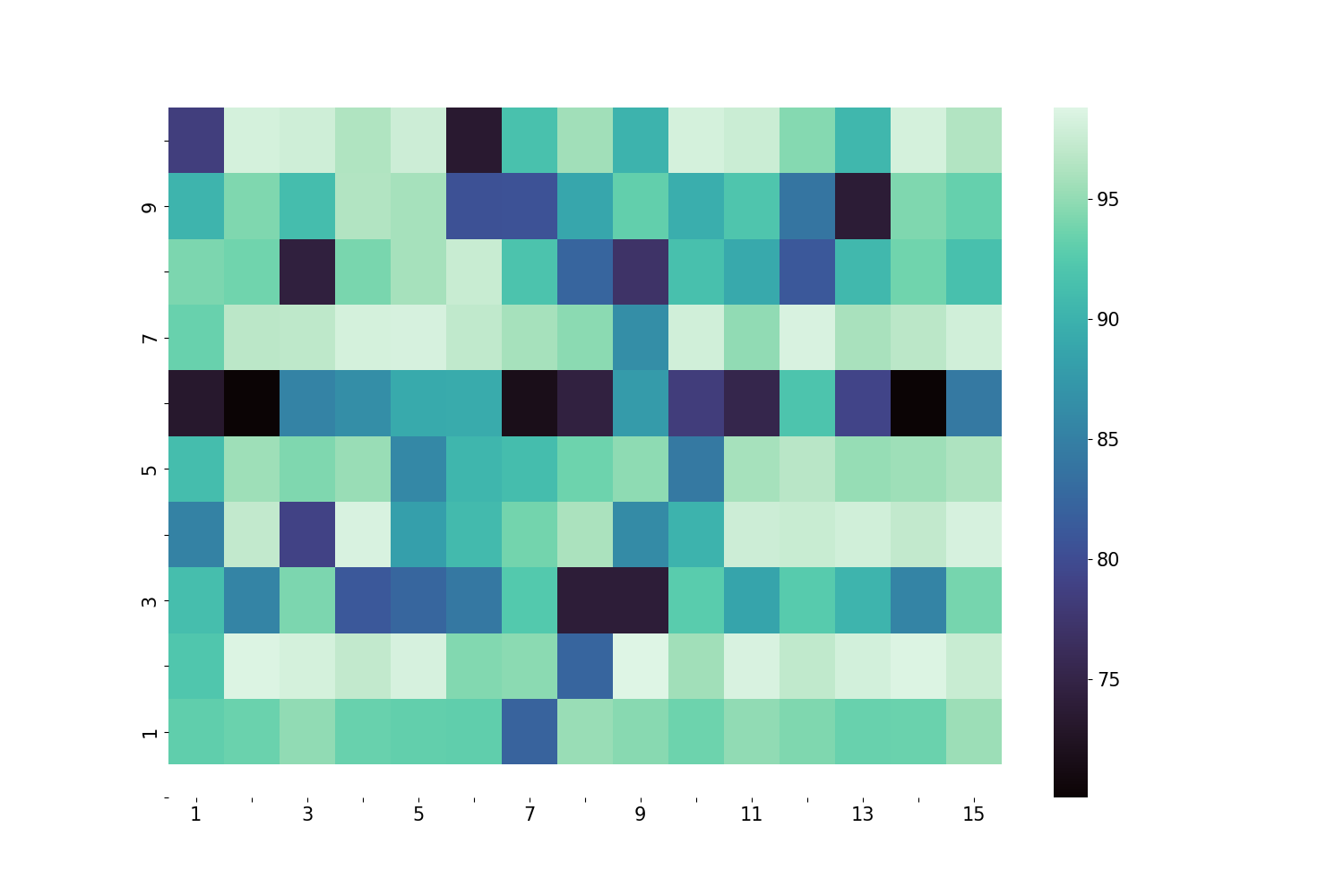}
  \caption{\scriptsize{BERT}\\r$_s.$ = 27.48\\Avg. = 90.97}
  \end{subfigure}
  % \begin{subfigure}{0.28\linewidth}
  \begin{subfigure}{0.45\linewidth}
    \centering
    \includegraphics[width=\linewidth, trim={0 0  8.5cm 0}, clip]{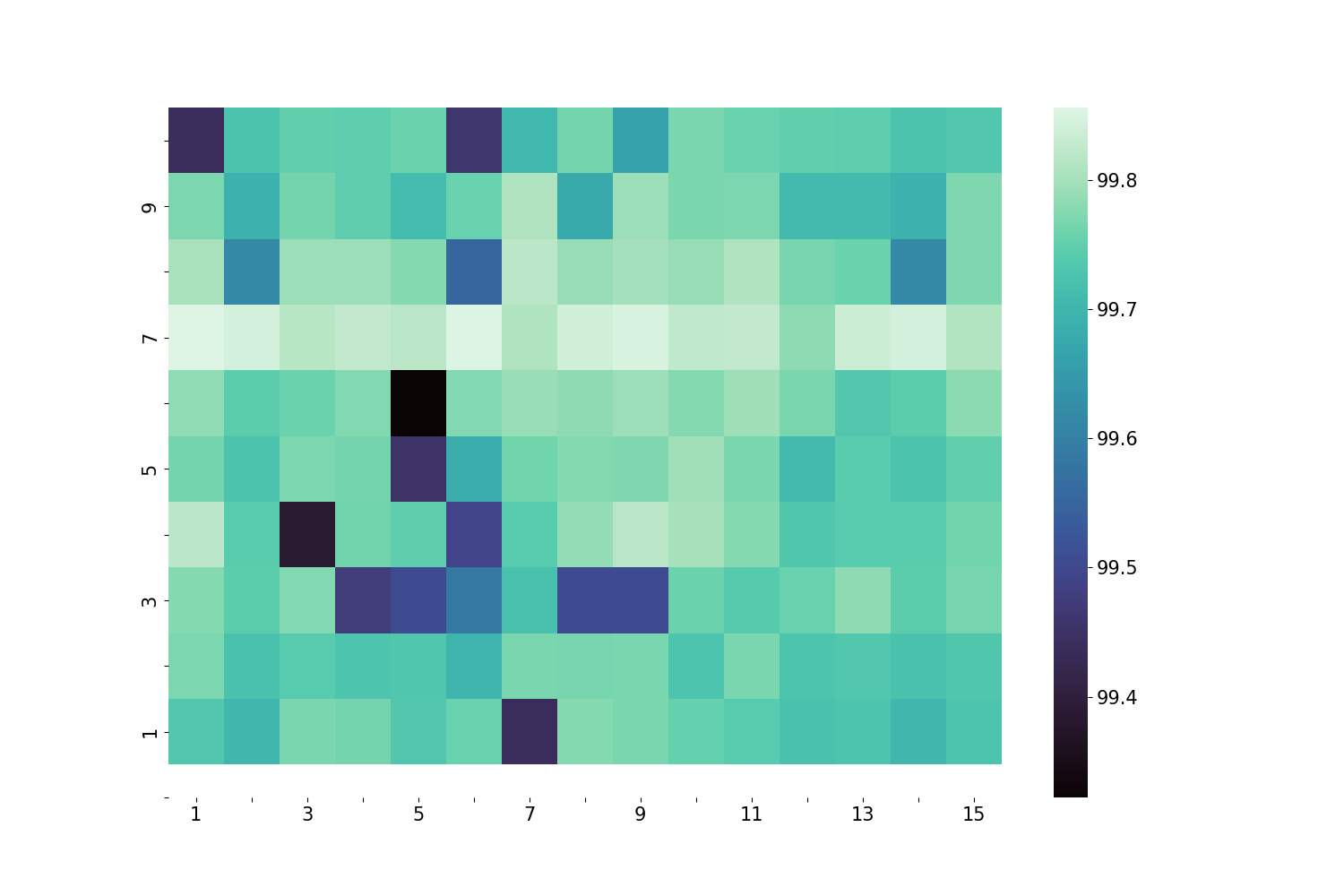}
    \caption{\scriptsize{RoBERTa}\\ r$_s.$ = 44.82\\Avg. = 99.74}
  \end{subfigure}
  % \begin{subfigure}{0.28\linewidth}
  \begin{subfigure}{0.45\linewidth}
    \centering
    \includegraphics[width=\linewidth, trim={0 0  8.5cm 0}, clip]{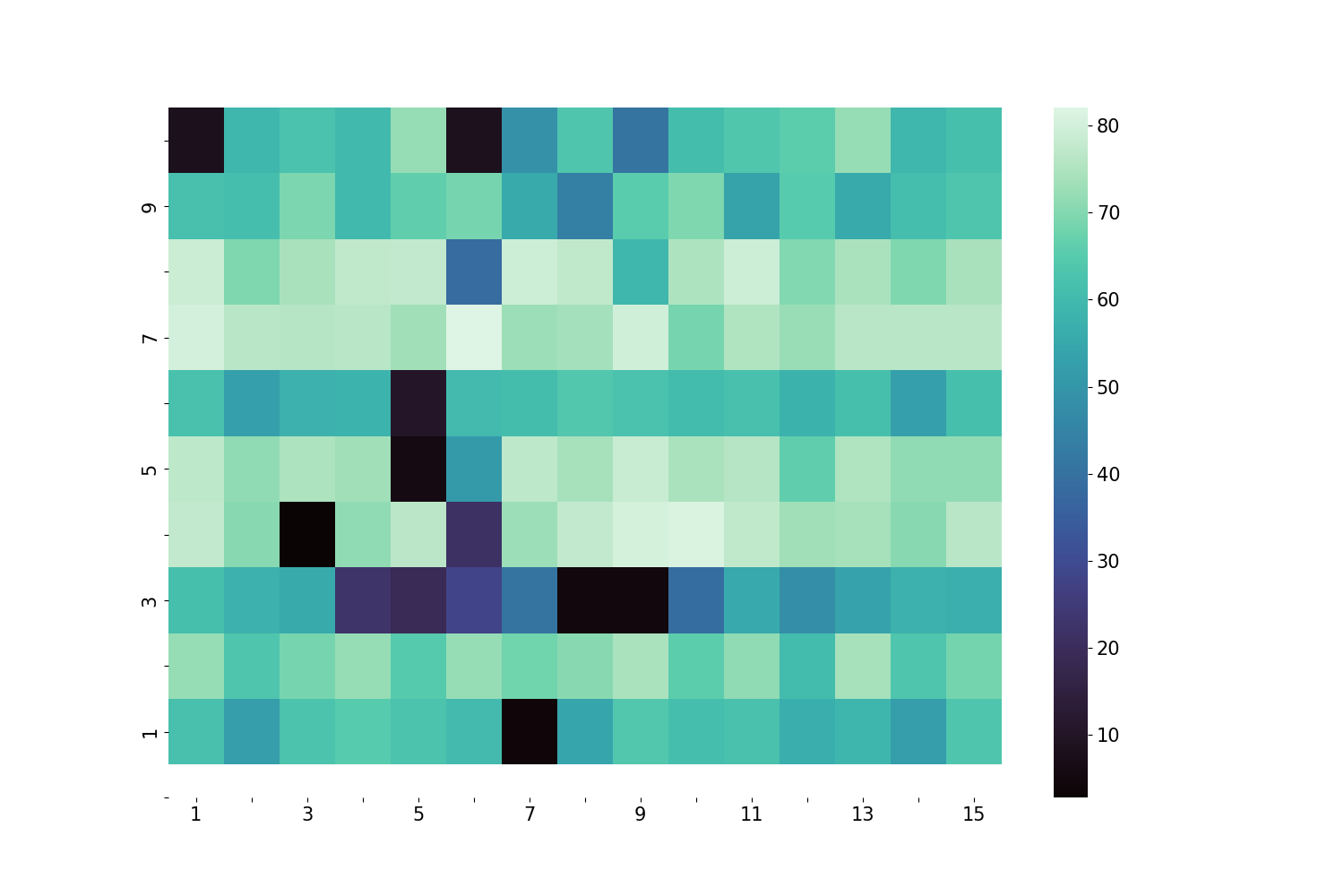}
    \caption{\scriptsize{OpenAI model}\\ r$_s.$ = 47.23\\Avg. = 61.79}
  \end{subfigure}
  \begin{subfigure}{0.15\linewidth}
    \centering
    \includegraphics[trim={28.5cm 0  0 0}, clip, height=3\linewidth]{fig/cos/gpt4v_prompt_cos_all_heatmap_openai_sort.png}
    \caption*{}
  \end{subfigure}
  \caption{COS$_F$ scores are presented in heatmaps, where the X-labels indicate different instructions and Y-labels different samples. The Spearman’s correlation coefficients are denoted as r$_s.$, with $p$-value $<$ 0.1\% in all cases, while Avg. indicates the average COS$_F$ score (notice that the OpenAI model is the one with the highest correlation while having at the same time a lower average cosine score. The other two models seem to have a high level of anisotropy of the vector space, cf. \citet{ethayarajh2019contextual,feng2025learning}, assigning similarities close to 1 to most text pairs). 
  Regarding model versions, we chose \textit{bert-base-uncased} for BERT, \textit{rflike berta-base} for RoBERTa, and \textit{text-embedding-3-small} for the OpenAI model. All scores are reported in percentage.}
  \label{fig:cos-heatmap} 
  \end{figure*}

 Following \citet{li2020sentence}, we designed a similarity experiment to find the ideal embedding model: 
\begin{itemize}
  \item First, we collected generated texts from GPT-4V using randomly-picked samples and designed instructions, and then created gold label similarities between pairs of gold truth captions and generated texts (The gold label similarity is actually a pseudo similarity. Currently, we haven't conducted a standardized human evaluation of the generated texts. But GPT-4V sometimes produces invalid responses due to sensitivity concerns or other reasons, so we take the extreme values by assigning a similarity score of 0 between the caption and invalid response pair, and a score of 1 for other cases.); 
  \item We obtained text embeddings from each candidate embedding model; 
  \item We computed COS$_F$ scores between each paired embeddings as the model-wise predicted similarities, and then calculate the Spearman’s correlation coefficients between them. We experimented with BERT, RoBERTA and the OpenAI embedding models, and reported the COS$_F$ and Spearman scores in Figure \ref{fig:cos-heatmap}. An ideal model is supposed to exhibit strong correlation with the gold labels. Additionally, it should be able to differentiate between different instructions, meaning that the predicted similarities should vary upon the provided instructions. Based on these criteria, the OpenAI embedding model outperformed the others. Besides, considering its larger token window (1536 tokens), we decided to select the OpenAI embedding model as the embedding source for the COS$_F$ metric.
\end{itemize}
% (1) first, we collected generated texts from GPT-4V using randomly-picked samples and designed instructions, and then created gold label similarities between pairs of gold truth captions and generated texts (The gold label similarity is actually a pseudo similarity. Currently, we haven't conducted a standardized human evaluation of the generated texts. But GPT-4V sometimes produces invalid responses due to sensitivity concerns or other reasons, so we take the extreme values by assigning a similarity score of 0 between the caption and invalid response pair, and a score of 1 for other cases.); 
% (2) we obtained text embeddings from each candidate embedding model; 
% (3) we computed COS$_F$ scores between each paired embeddings as the model-wise predicted similarities, and then calculate the Spearman’s correlation coefficients between them. We experimented with BERT, RoBERTA and the OpenAI embedding models, and reported the COS$_F$ and Spearman scores in Figure \ref{fig:cos-heatmap}. An ideal model is supposed to exhibit strong correlation with the gold labels. Additionally, it should be able to differentiate between different instructions, meaning that the predicted similarities should vary upon the provided instructions. Based on these criteria, the OpenAI embedding model outperformed the others. Besides, considering its larger token window (1536 tokens), we decided to select the OpenAI embedding model as the embedding source for the COS$_F$ metric.
\section{Supplementary Tables and Figures}\label{apd:first}
\begin{table*}[tb!]
\centering
\small
    \begin{tabular}{p{0.35\linewidth} p{0.2\linewidth} p{0.1\linewidth} p{0.05\linewidth} p{0.15\linewidth}}
    \hline
    \textbf{Dataset} & \textbf{Domain} &  \textbf{Total} &  \textbf{Caps.\footnotemark{}} &   \textbf{Source} \\
    \hline
    Conceptual Caption \cite{sharma2018conceptual} & Generic & 3.3M & 5 & Web \\ % ghandi, captions = 1
    MS COCO  \cite{lin2014microsoft} & Generic & 328K & 5 & Web \\  %??， 328K ？ 132K?
    Flickr30k \cite{young2014image} & Generic (people, animals) & 31K & 5 & Flickr.com \\
    Flick 8K \cite{hodosh2013framing} & Generic & 8K & 1-5 & Flickr.com  \\
    FlickrStyle10K \cite{gan2017stylenet}  & Generic  & 10K & 2 & Flickr.com\\
    SBU Captions \cite{ordonez2011im2text} & Generic & 1M & 5 & Web \\
    Visual Genome \cite{krishna2017visual} & Generic & 108K & & Web \\  % 5.4M R.D. region description
    VizWiz Captions \cite{gurari2020captioning} & Assistive & 39K & 5 & VizWiz APP \\  % ghandim 39K 
    CUB-200 \cite{welinder2010caltech} & Birds & 12K & 10 & Web \\
    Oxford-102 \cite{nilsback2008automated} & Flowers & 8K & 10 & Web \\
    Fashion Captions \cite{yang2020fashion} & Fashion & 52K & 5 & Web \\
    BreakingNews \cite{ramisa2017breakingnews} & News(sports, arts, etc.)  & 100K & 5 & Web \\
    GoodNews \cite{biten2019good} & News & 466K & 1 & New York Times\\% API\\
    SentiCap \cite{mathew2022infographicvqa} & Generic & 3.2K & 6 & MS COCO\\ % 
    TextCaps \cite{sidorov2020textcaps} & OCR %(reading comprehension) 
    & 28.4K & 5-6 & Web\\% (Open Images v3)\\ 
    nocaps \cite{agrawal2019nocaps} & Generic & 15.1K & 11 & Web \\%, Open Images V4 \\
    % sharm
    IAPR TC-12 \cite{grubinger2006iapr}  & Generic & 20K & 1-5 & Viventura \\%(a travel company)  \\
    %  (sports, actions, people, animals, cities, landscapes, daily life) 
    PASCAL 1K \cite{rashtchian2010collecting}& Generic (people, animals)& 1K & 5 & PASCAL VOC \\
    % \hline
    % Instagram dataset (for post generation) & Generic (design, food, style, etc.) & 721K  & & Instagram API\\ % people     
    \hline
    \end{tabular}
    \caption{A general summary of the benchmark datasets for image captioning}
    \label{tab:datasets_review}
\end{table*}
\footnotetext{The number of captions per image.}
% % err case
\begin{figure*}[htb!]
  \centering
  \begin{subfigure}{0.28\linewidth}
      \includegraphics[width=0.9\textwidth, trim={0 0 6.8cm 0}, clip]{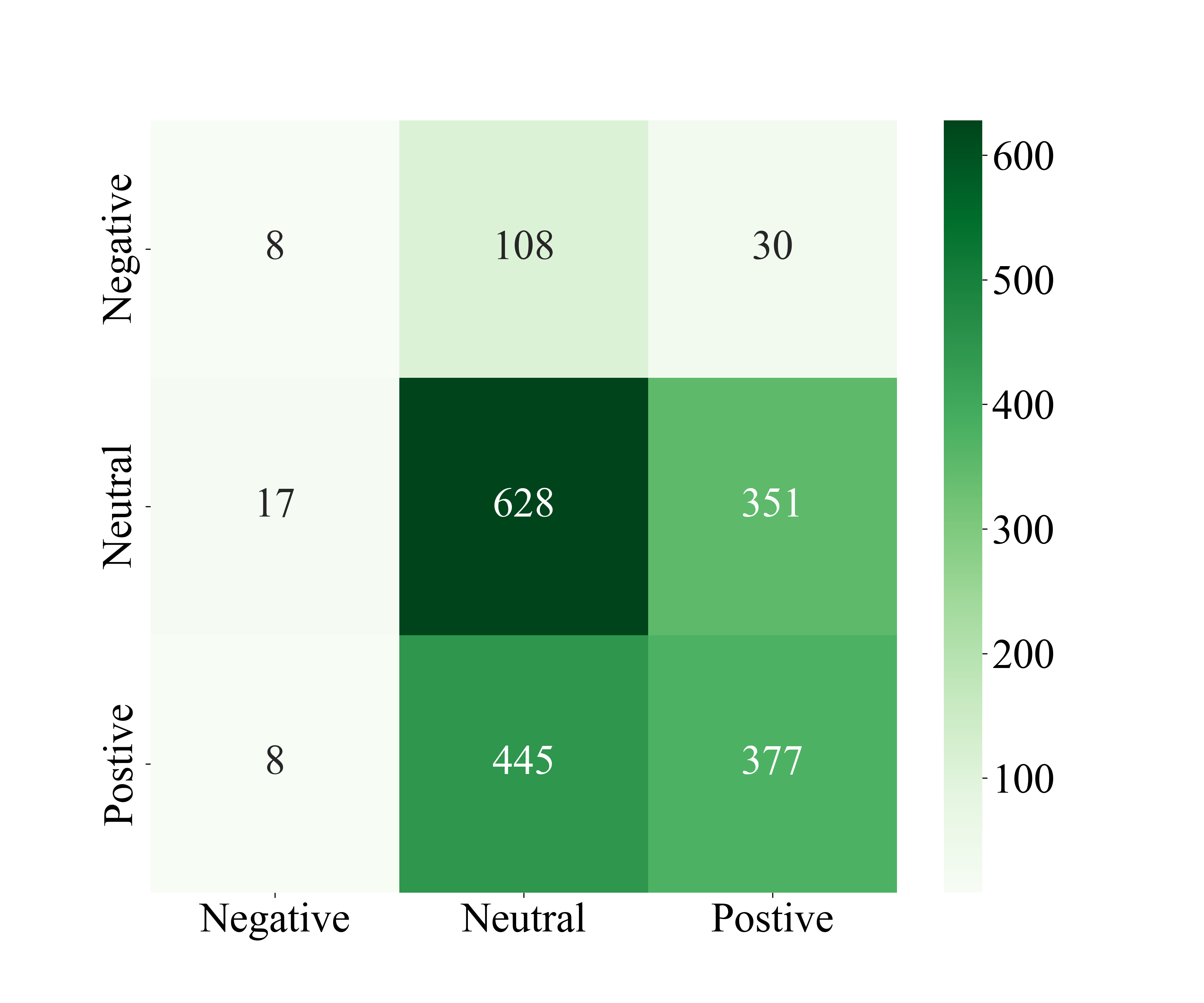}
      \caption{GPT-4V \\\centering(F$_1$ =38.04)}
  \end{subfigure}
  \begin{subfigure}{0.28\linewidth}
      \includegraphics[width=0.9\textwidth, trim={0 0 6.8cm 0}, clip]{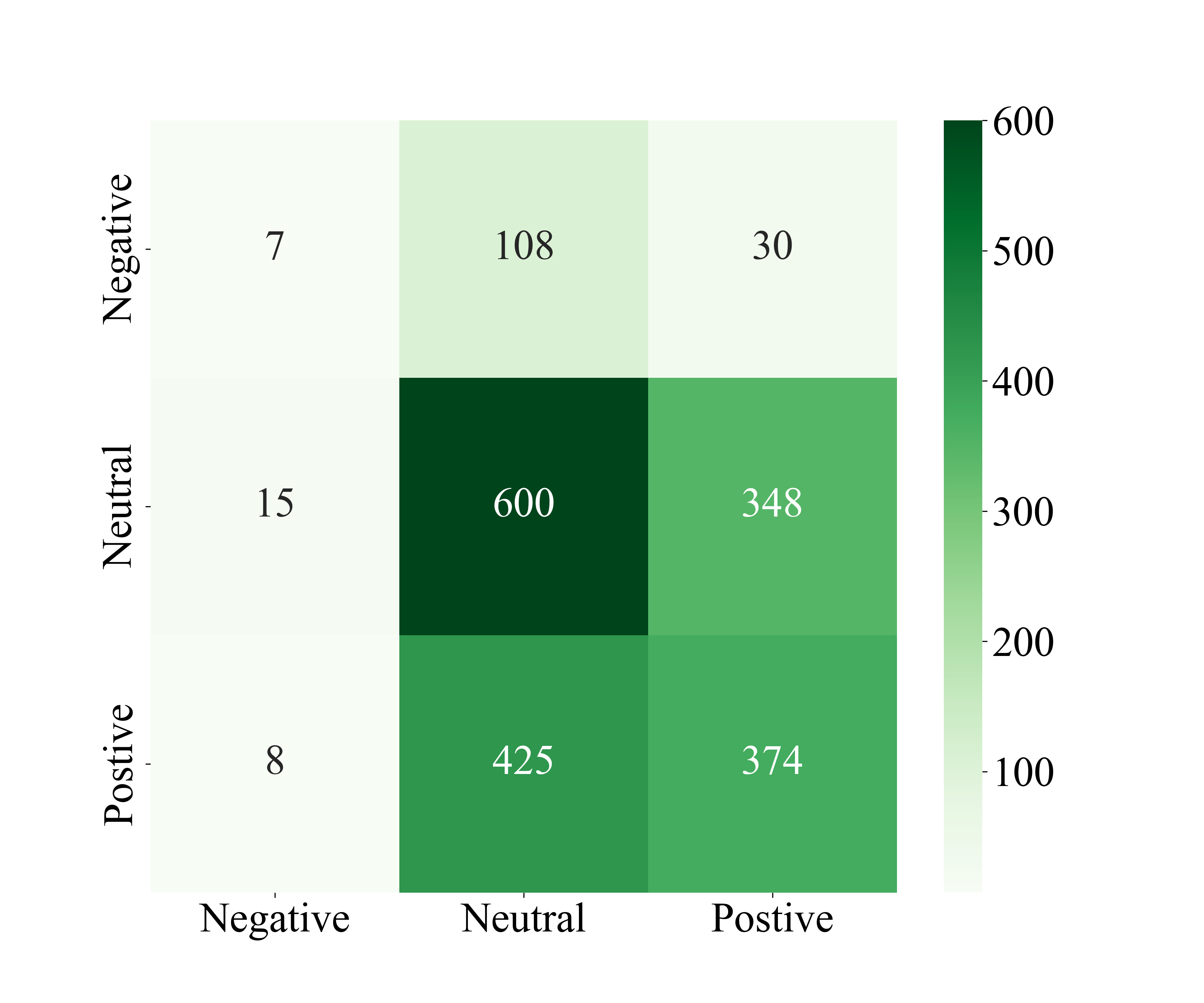}
      \caption{GPT-4V (Valid)\\\centering(F$_1$ =37.74)}
  \end{subfigure}
  \begin{subfigure}{0.28\linewidth}
      \includegraphics[width=0.9\textwidth, trim={0 0 6.8cm 0}, clip]{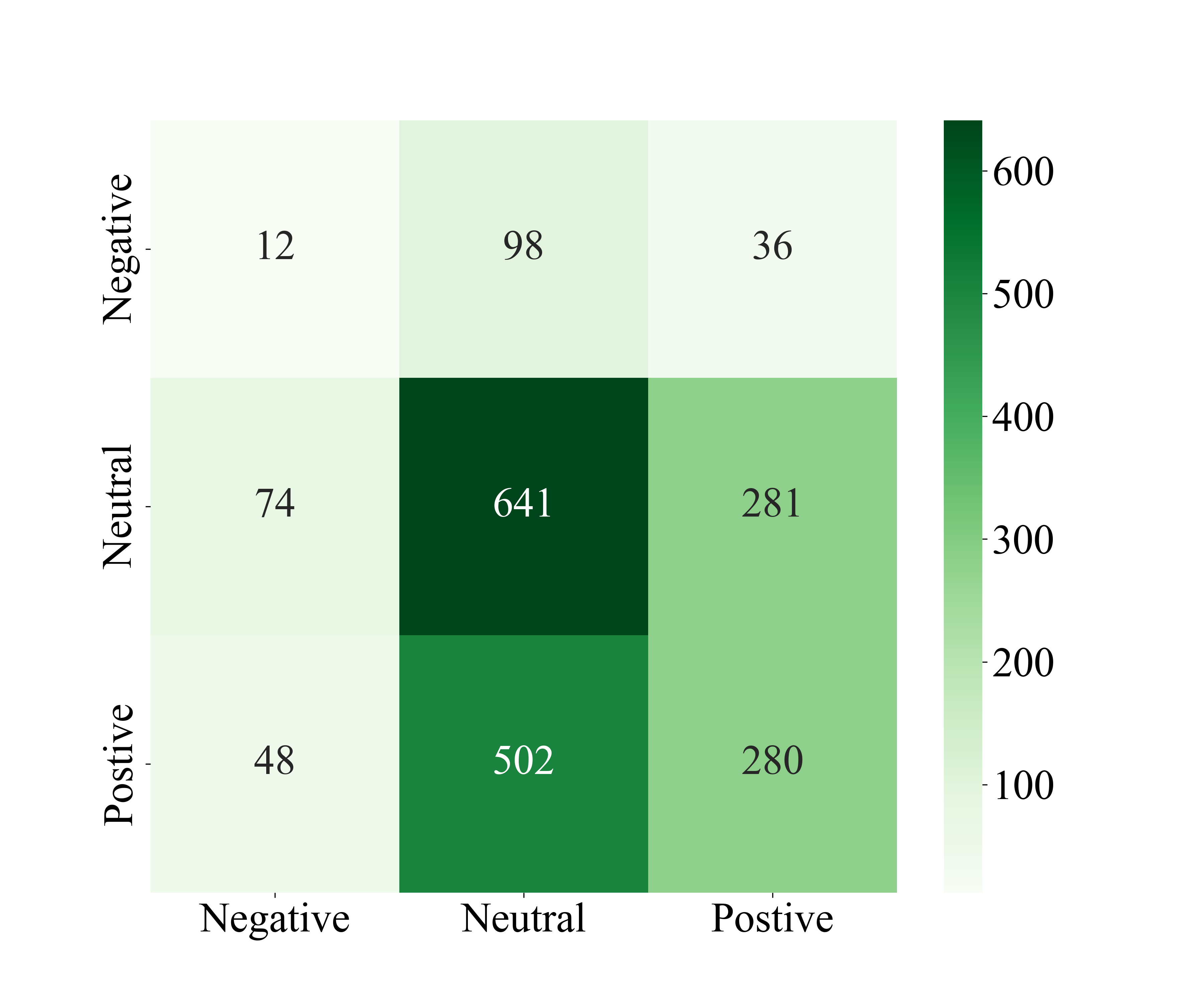}
      \caption{mPLUG-Owl2 \\\centering(F$_1$ =35.04)}
  \end{subfigure}
  \begin{subfigure}{0.28\linewidth}
      \includegraphics[width=0.9\textwidth, trim={0 0 6.8cm 0}, clip]{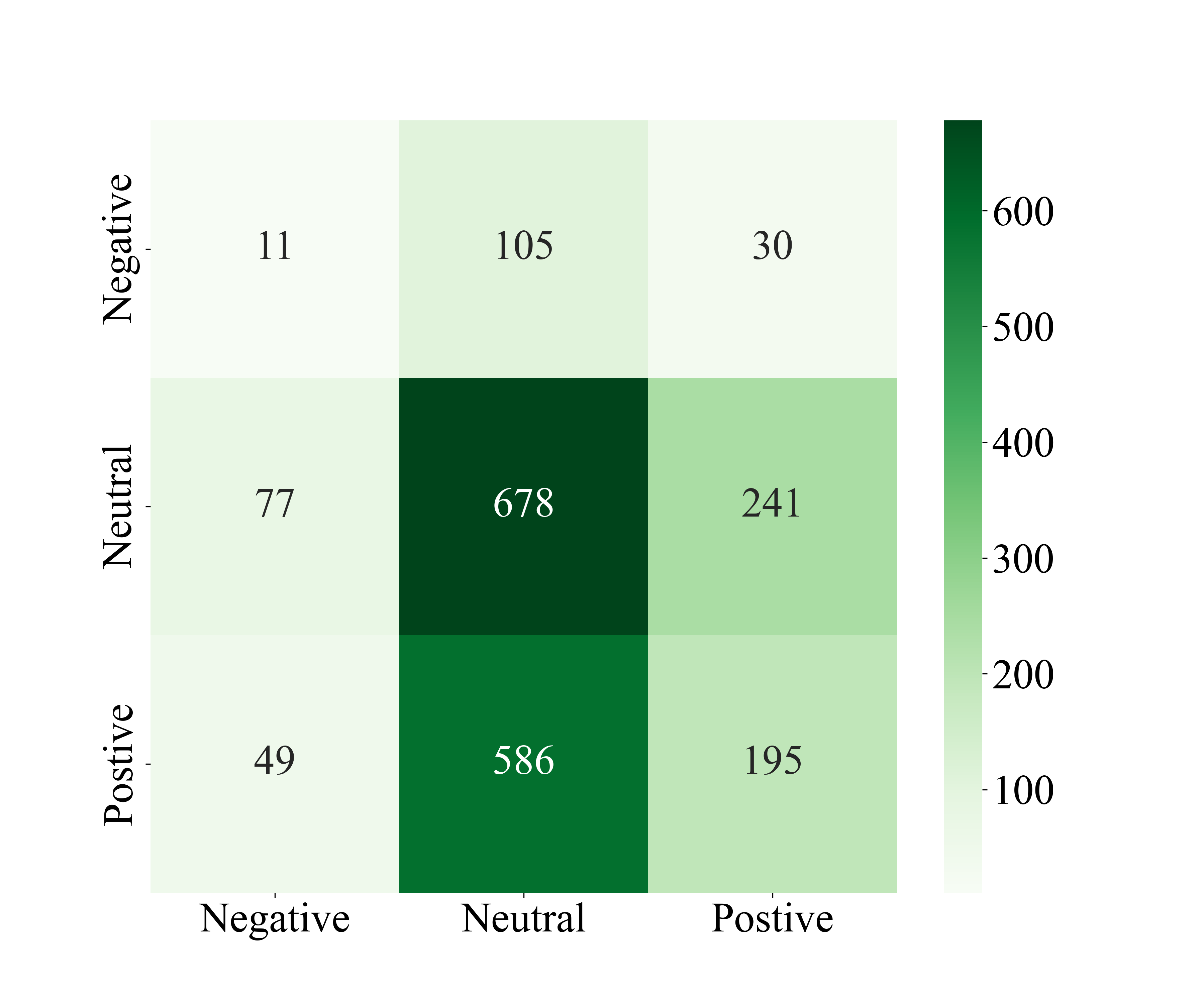}
      \caption{LLaVA \\\centering(F$_1$ =31.73)}
  \end{subfigure}
  \begin{subfigure}{0.28\linewidth}
      \includegraphics[width=0.9\textwidth, trim={0 0 6.8cm 0}, clip]{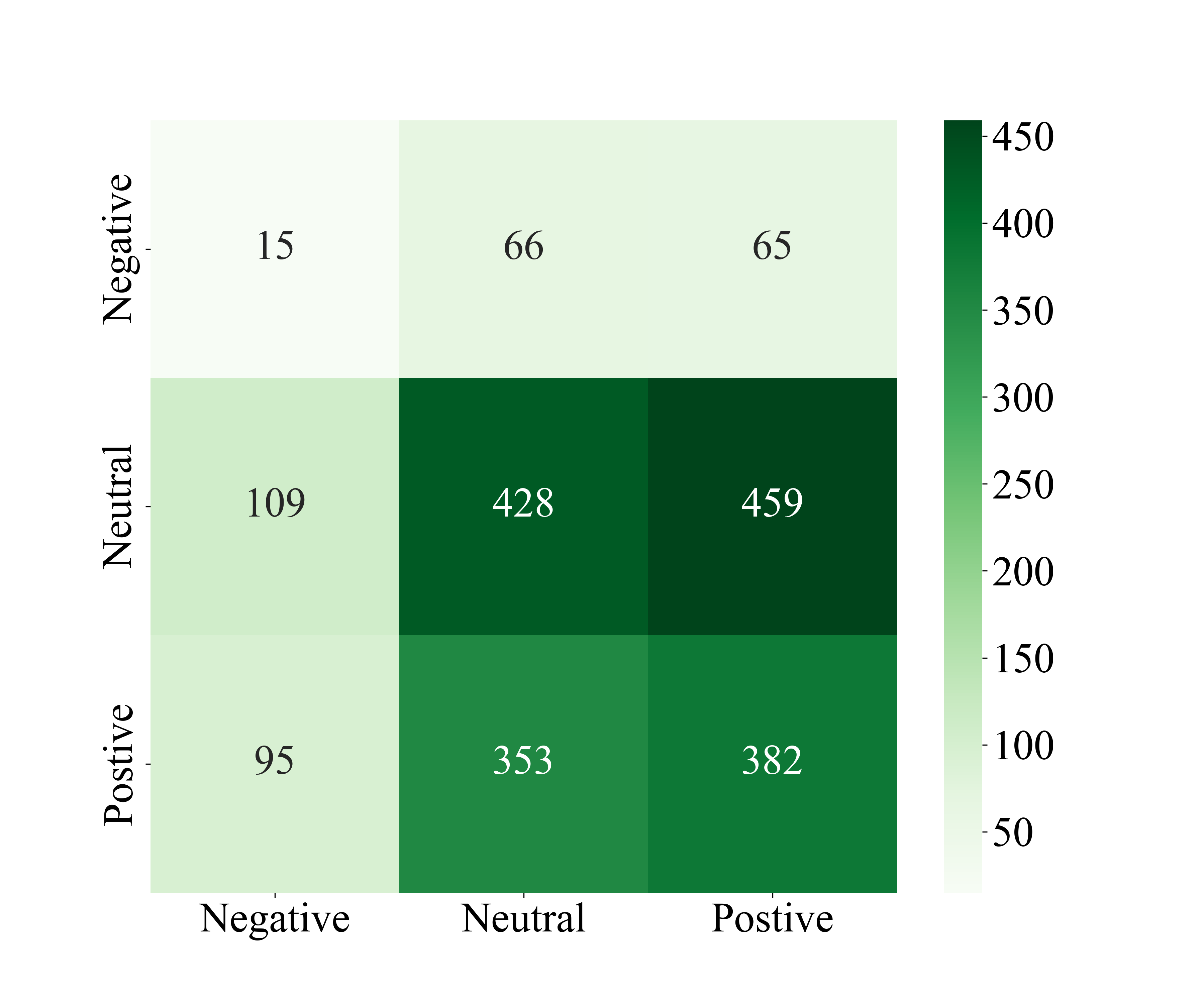}
      \caption{InstructBLIP \\\centering(F$_1$ =32.89)}
  \end{subfigure}
  \begin{subfigure}{0.28\linewidth}
      \includegraphics[width=0.9\textwidth, trim={0 0 6.8cm 0}, clip]{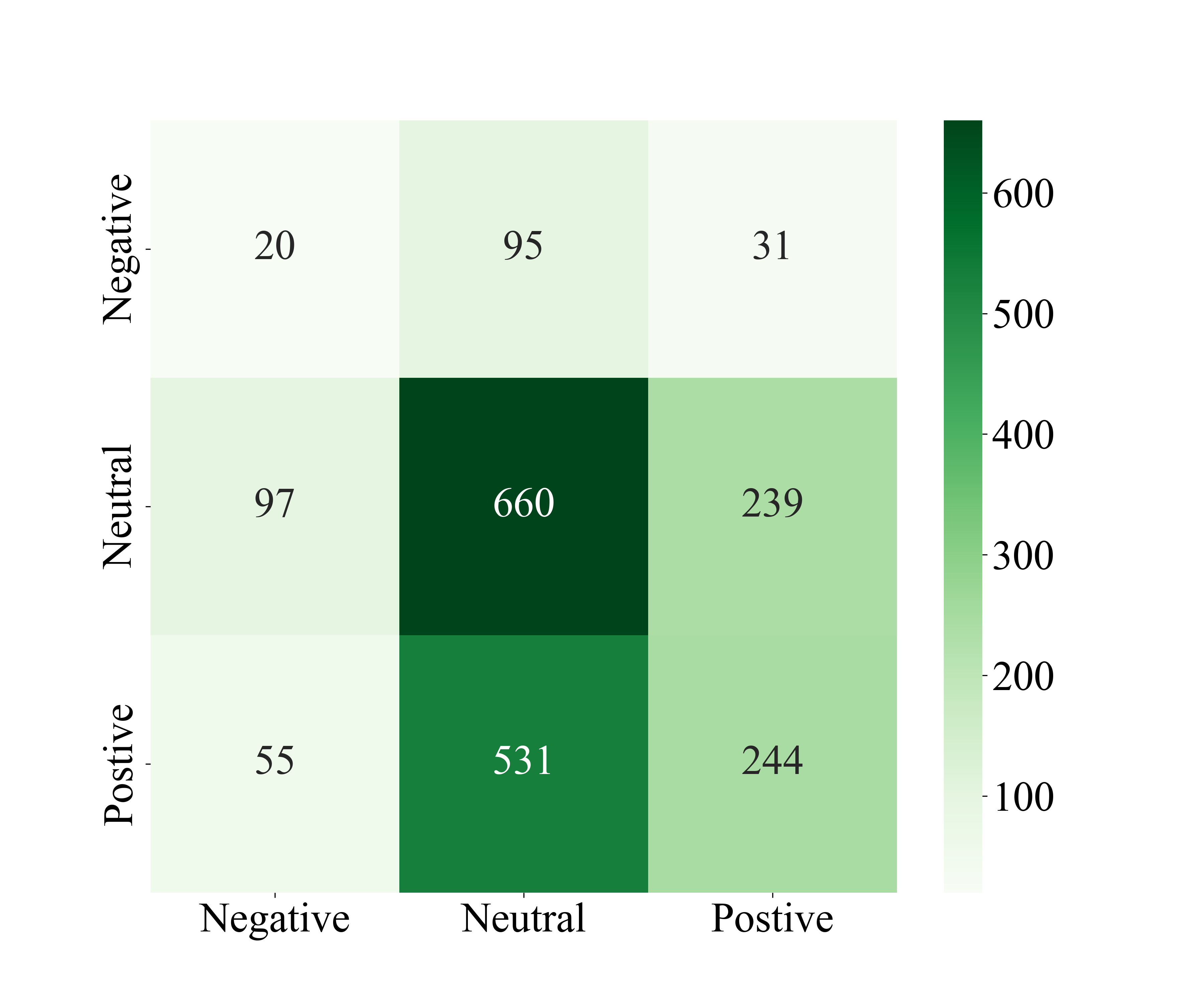}
      \caption{Gemini \\\centering(F$_1$ =35.58)}
  \end{subfigure}
  \begin{subfigure}{0.1\linewidth}
      \includegraphics[width=0.8\textwidth, trim={24cm 0 0 0}, clip]{fig/case_study/gemini_matrix.png}
  \end{subfigure}
  \caption{Confusion matrices of SA. Macro F$_1$ scores of the models are presented in the brackets, reported on a 100\% scale.}
  \label{fig:matrix_sa}
\end{figure*}
% % err case
\begin{table*}[hpt!]
  \centering
  \caption{An example of the false neutral case from Gemini (i.e., a positive sample is classified as a neutral). Only the last a few sentences are presented.}
  \label{tab:sa_case}
  \begin{tabular}{l p{0.9\linewidth}}
  \multicolumn{2}{c}{\includegraphics[width=0.85\linewidth]{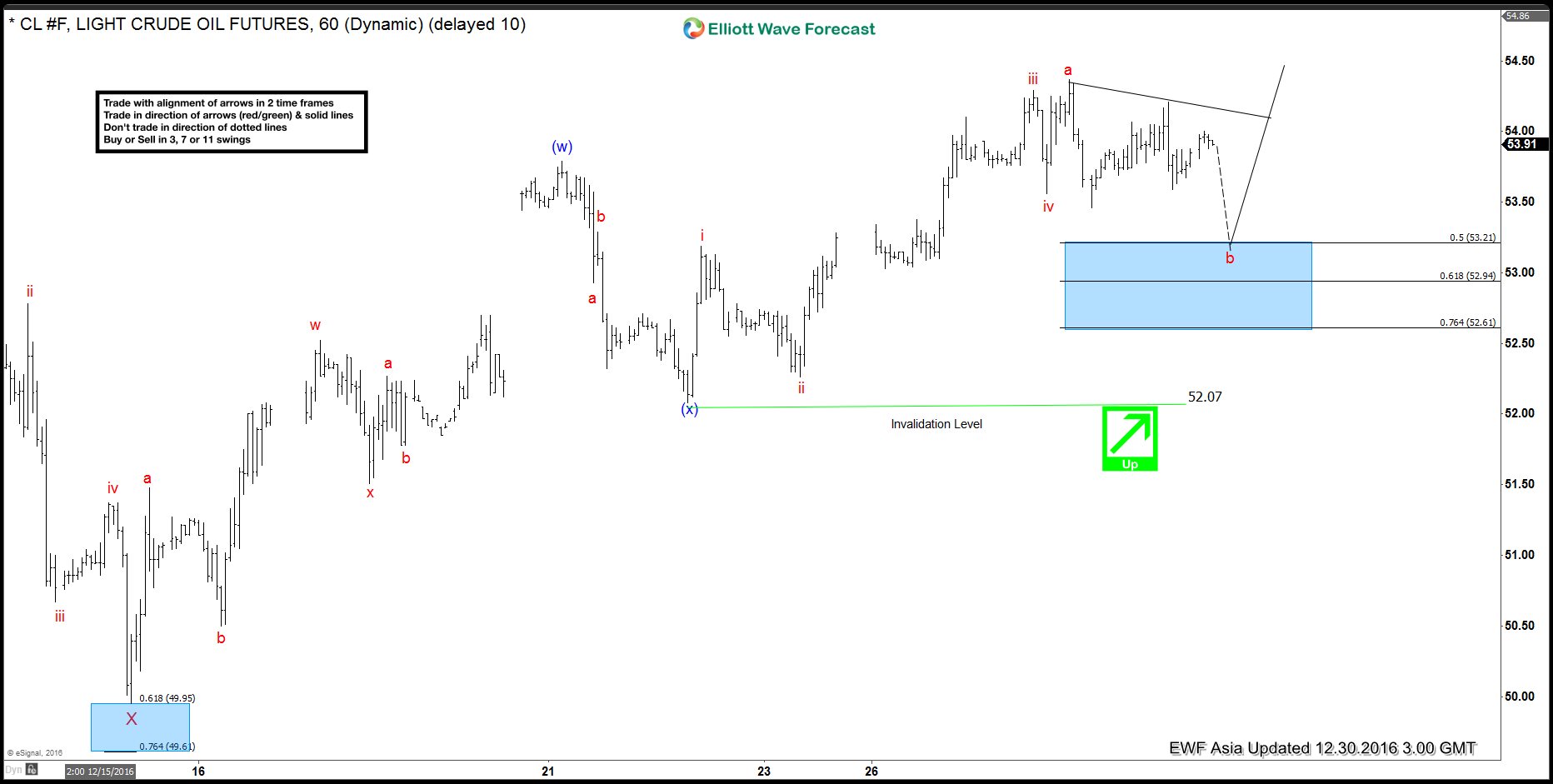}}\\
  \\
\textbf{Reference}: & "... The internal structure of wave Z is unfolding as a double three structure where wave (w) of Z ended at  \$53.79 and wave (x) of Z has ended at \$52.08. Near term, the rally from wave (x) low ended at \$54.37 as 5 waves, and CL\_F is expected to do a wave b pullback to correct the rally from \$52.08 low in 3, 7, or 11 swing before the rally resumes towards \$55.93 – \$56.83. We don’t like selling the proposed pullback and expect buyers to appear again once wave b pullback is complete in 3, 7, or 11 swing as far as pivot at \$52.07 remains intact in the first degree. If pivot at \$52.07 fails, then CL\_F has ended cycle from \$49.95 low and will do a larger correction and still expected to turn higher while second degree pivot at \$49.95 stays intact."\\  % sent = pos
  \\     
\textbf{Gemini}: & 
  ... The identified wave structure combined with critical support levels provides a framework for possible upward momentum. Traders and investors are advised to watch for a rebound from the defined support levels, which could signify a continuation of the bullish trend established earlier in the cycle. This analysis reinforces a strategic approach, encouraging trades in alignment with the identified Elliott wave patterns and Fibonacci levels for optimal risk management.\\  % sent = neutral
  \end{tabular}
\end{table*}

\begin{table*}[hpt!]
    \scriptsize
    \centering
    \begin{tabular}{p{0.08\linewidth} p{0.9\linewidth}}
    \multicolumn{2}{c}{\includegraphics[width=0.8\linewidth]{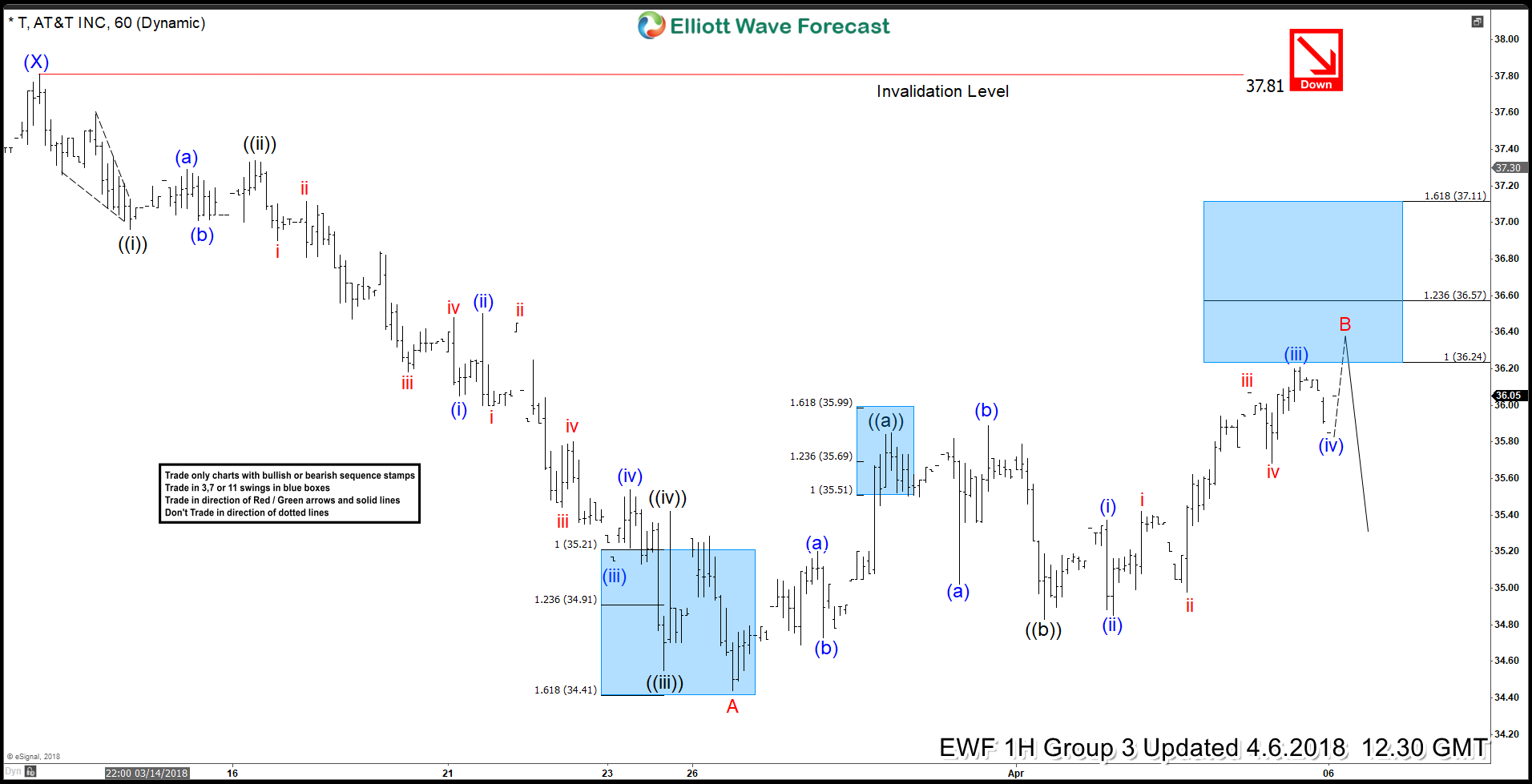}}\\
    \\
    Reference: & AT\&T ticker symbol: (\$T) short-term AT\&T Elliott Wave view suggests that the bounce to 3\/13 high (37.81) ended Intermediate wave (X) bounce. Below from there, intermediate Y leg lower remains in progress as a Zigzag correction. Where initial decline unfolded in 5 waves Elliott Wave Impulse Sequence with extension in Minor wave A. Which can be seen in the Chart below, with the internal distribution of each leg consisted of 5 waves structure thus it favored being impulse. Now let’s take a quick look at the short-term price action in the stock from 3\/13 peak (37.81). The decline from there can be seen in 5 waves impulse sequence, where internals of each leg is consist of 5 waves structure. Where Minute wave ((i)) ended at 36.97, Minute wave ((ii)) ended at 37.34 and Minute wave ((iii)) ended at 34.55. Then from there, Minute wave ((iv)) ended at 35.29 and Minute wave ((v)) of A ended at 34.44 low. Up from there, the stock is correcting the 3\/13 cycle (37.81) in Minor wave B bounce. And expected to find seller’s in 3, 7 or 11 swings as far as the pivot from 3\/13 peak (37.81) stays intact. So far the bounce looks to be in 3 swings as a Flat structure, where Minute wave ((a)) ended at 35.89, Minute wave ((b)) ended at 34.84 and Minute wave ((c)) of B remains in progress towards 36.24-36.57 100\%-123.6\% Fibonacci extension area of ((a))-((b)). Before finding seller’s again for another extension lower ideally provided the pivot from 37.81 high remains intact. We don’t like buying it and intraday traders can look to sell it from an above-mentioned 100\%-123.6\% ext area for 3 wave reaction lower at least. Keep in mind that the market is dynamic and the FTSE Elliott Wave view could change in the meantime...\\\\
    Gemini: & This Elliott Wave Forecast chart for AT\&T INC (60-minute timeframe) depicts historical price movements and projects future trends using Elliott Wave principles. The chart shows a complex pattern, marked by distinct impulsive and corrective waves labeled with Roman and lowercase letters respectively. The analysis identifies two completed corrective patterns labeled 'A' and 'B'. 'A' shows a corrective three-wave structure, followed by an impulsive five-wave structure labeled 'B'. The 'B' wave appears to have reached a potential termination point, suggesting a significant price reversal is imminent. A dashed line indicates a potential invalidation level at approximately 37.81. If the price breaks above this level, the bearish projection would be invalidated. The chart's creator has drawn blue boxes around specific wave structures, suggesting these areas represent significant price swings suitable for trading. Red and green arrows indicate the predicted direction of price movement; dotted lines, however, should be avoided for trading signals. The analyst recommends trading only charts displaying clear bullish or bearish sequence stamps and focusing on 3, 7, or 11 swings within the blue box areas. The chartist's overall opinion is bearish. The structure of wave 'B' and its potential termination point suggest that a downward trend is likely to follow. The invalidation level at 37.81 provides a clear benchmark for assessing the validity of this bearish prediction. A break above this level would signal a significant shift in the underlying price trend and invalidate the current bearish outlook. The analyst emphasizes the importance of adhering to the specified trading rules, particularly avoiding trades counter to the direction of the solid and directional arrows.\\\\
    GPT-4V: & The chart displaying the historical price movements of AT\&T Inc. (Ticker: T) utilizes the Elliott Wave theory to forecast future trends based on observed patterns in the price movements. The analysis reveals the stock has experienced significant volatility, characterized by both upward and downward movements delineated by specific wave formations. Initially, the stock faced a bearish trend, denoted by the downward movements labeled as waves (i), (ii), and (iii). This initial decline appears to have established a low point around 34.41, suggesting potential support levels. Subsequently, the price exhibited a corrective wave structure wherein it managed to regain some value, as indicated by the formations tagged as ((iii)), ((iv)), and subsequent retracements represented by (a) and (b). The chart depicts key Fibonacci retracement levels, illustrating potential areas of resistance and support. For instance, the (1) level at 35.51 indicates a notable area where the price experienced fluctuations, while the 1.236 and 1.618 levels offer insights into possible price targets for the next wave movements. The areas shaded in blue emphasize critical zones for potential trading decisions, guiding market participants on optimal entry and exit points. Moving forward, the chartist's...\\\\
    LLAVA: & The chart image displays a historical price movement of a stock, with a red line indicating a downward trend. The stock has been experiencing a decline \textcolor{red}{since the beginning of the year, with a notable drop in January}. The chartist's opinion on the stock trends is that the stock is likely to continue its downward trajectory, as the current trend is not showing any signs of reversal. The chart also features a forecast for the stock's future performance, with a red line indicating a potential decline. The chartist believes that the stock's current downward trend is likely to continue, and the forecast suggests that the stock may reach a lower low in the near future. The chartist's conclusion is that the stock is not a good investment at the moment, as the current trend is not favorable for long-term investors. The stock's historical price movements and the forecast indicate that it may continue to decline, making it a risky investment option.

    \end{tabular}
    \caption{We selected a chart image and presented it with its reference description and several model output for comparison. The chart presents a 1-hour short term movement of AT\&T ticker. } 
    \label{tab:text_comp}
\end{table*}
\end{document}